\documentclass[paper=a4, fontsize=11pt]{scrartcl} 
\pdfoutput=1 

\usepackage[T1]{fontenc} 
\usepackage{color}
\usepackage[english]{babel} 
\usepackage{amsmath,amsfonts,amsthm} 
\usepackage{graphicx}
\usepackage{hyperref}
\usepackage{amssymb}

\begin{document}

\begin{titlepage}
\begin{center}

{\Large \textbf{ A simple model for the evolution of a non-Abelian cosmic string 
	network}}

 ~\\
\vskip 2cm

\vskip 1cm

{  \textbf{G. Cella$^{a}$\footnote{ giancarlo.cella@pi.infn.it},   \textbf{M. Pieroni$^{bc}$ }\footnote{\textbf{Corresponding author}: mauro.pieroni@apc.univ-paris7.fr} }}\\
~\\
~\\
{\em ${}^a$ Istituto Nazionale di Fisica Nucleare, sez. Pisa\\ Largo Bruno Pontecorvo 3, 56126 Pisa} \\
{\em ${}^b$ AstroParticule et Cosmologie,Universit\'e Paris Diderot, CNRS, CEA\\Observatoire de Paris, Sorbonne Paris Cit\'e, F-75205 Paris Cedex 13} \\
{\em ${}^c$ Paris Centre for Cosmological Physics\\F-75205 Paris Cedex 13}\\

\end{center}

\vskip 1cm
\centerline{ {\textbf{ Abstract}}} {In this paper we present the results of numerical simulations intended to study 
	the behavior of non-Abelian cosmic strings networks. In 
	particular we are interested in discussing the variations in the asymptotic 
	behavior of the system as we variate the number of generators for the 
	topological defects. A simple model which allows 
	for cosmic strings is presented and its lattice discretization is discussed. 
	The evolution of the generated cosmic string networks is then studied for 
	different values for the number of generators for the topological defects. 
	Scaling solution appears to be approached in most cases and we present an 
	argument to justify the lack of scaling for the residual cases.}

\indent

\vfill

\vfill

\end{titlepage}
\date{\normalsize\today} 

\tableofcontents

\section{\label{sec:Introduction}Introduction}

The notion of spontaneous symmetry breaking (SSB) is a main concept of modern 
theoretical physics and its applications are indeed widespread in all physical 
areas. Under some particular conditions this mechanism may lead to the formation 
of topological defects such as strings, monopoles and domain 
walls~\cite{HindmarshKibble1995,cstd}.
It is known that only the first among these objects may have
some relevance in a cosmological context~\cite{NODOMINI,nomonop}. In particular SSB at Grand Unification Scales (GUT) or lower may generate cosmic strings with $G \mu \lesssim 10^{-6}$, where $G$ is the Newton constant and $\mu$ is the string tension. It is interesting to point out that the presence of these objects in our universe may generate some observational signature. In particular, inspired by the works of Berezinsky, Hnatyk and Vilenkin~\cite{Berezinsky:2000vn,Berezinsky:2001cp}, Damour and Vilenkin have shown in~\cite{Damour:2000wa,DamourVil2,DamourVil1} that some GW emissions from cosmic strings may be observed via pulsar timing observations\cite{Lommen:2015gbz} or by higher frequency experiments such as LIGO/VIRGO interferometers~\cite{LIGO,VIRGO} or the eLISA mission~\cite{Vitale2014}. Cosmic strings are widely discussed in literature~\cite{Vachasp+Vilenk,Hindmarsh:1988jn,KibbleTurok1982} and it is well known that in the zero-width limit, their dynamics is well approximated by Nambu action. On the 
contrary modelling a network of cosmic strings presents non trivial issues and 
numerical simulations are required to have a proper comprehension 
of the behavior of the system.

Well known results on networks of Abelian cosmic strings in a Friedmann-Lema\^itre-Robertson-Walker (FLRW) background have been derived by Albrecht and Turok in~\cite{ALB+TUROK}. In this work the authors argue about the possibility of describing the system in terms of a single scale parameter. Under this assumption they show the appearance of a scaling solution that is not in contrast with experimental observations. Several generalizations of the one-scale models have been proposed, in particular it is worth mentioning the three scales model proposed by Austin, Copeland and Kibble in~\cite{AustinCopelandKibble1993}. Another interesting proposal has been formulated by Martins and Shellard in~\cite{MartinsShellard1996}, who generalized the one-scale model to include the effects of frictional forces. In both of these generalized models scaling solutions still appears to be reached under quite general conditions. It is important to point out that the appearance of a scaling solution in this framework requires a mechanism that ensures an efficent chopping of long strings into smaller loops. This is usually ensured by the condition $p \sim 1$, where $p$ is the probabilty that string intercommute, i.e. exchange ends during a collision. Another assumption of the one scale description is that newly formed string loops have a typical size $l \sim \alpha t$ where the dimensionless constant $\alpha$ is usually called the loop-length parameter. In the standard picture the value of $\alpha$ is directly related to the gravitational damping of small-scale wiggles in particular we find $\alpha \sim \Gamma G \mu$, where $\Gamma \sim 50$ is a numerically determined coefficent \cite{cstd}.

Some critics to these simplified descriptions have been presented by Damour and Vilenkin in~\cite{DamourVil1} where they discuss some consequences of modifying the standard framework summarized in the above paragraph. This critical review is mainly focused on the two assumptions $p \sim 1$, $l \sim \alpha t$. As discussed by Bettencourt, Laguna and Matzner in~\cite{BettencourtLagunaMatzner1997}, it is actually possible to consider configurations\footnote{Type I non abelian cosmic strings with zippers.} where cosmic strings may not intercommute. In this case the standard $p \sim 1$ assumption can not be trusted anymore and this may have some interesting consequences affecting the evolution of the network. On the other hand, the works of 
Siemens Olum and Vilenkin~\cite{Siemens2,Siemens:2002dj}, show that the parametrization $l \sim \Gamma G \mu t$ can not be trusted. In particular, as discussed in~\cite{DamourVil1}, defining $\epsilon \equiv \alpha/\alpha^{st}$, where $l \sim \alpha t$ and $\alpha^{st} \sim \Gamma G \mu$, both the cases $\epsilon \ll 1$ and $\epsilon \gg 1$ need to be studied.   

In this paper we focus on the case of non-Abelian cosmic strings networks. To give a brief review of the state of art on this topic let us classify the existing numerical simulations into three categories:
\begin{itemize}
	\item We call \textbf{``Field-oriented''} simulations the numerical studies where the evolution is focused on the behavior of the Higgs field associated with the SSB. A first example of this class of simulations has been proposed by Spergel and Pen in~\cite{Spergel} and a more recent study of the same model has been proposed by Copeland and Saffin in~\cite{CopelandSaffin2005}. Another 
	interesting study of this class has been proposed by Hindmarsh and Saffin in~\cite{HindmarshSaffin2006}. 
	\item In a \textbf{``String-oriented''} simulation, once the initial conditions for the string network are generated, the evolution is directly described in terms of the string dynamics. The main references for this class of simulations are the works of McGraw~\cite{McGraw1,McGraw:1996py}.
	\item Finally we define \textbf{``Phenomenological''} simulations the class of works where the authors, motivated by phenomenological arguments, define a set of differential equations to describe the evolution of the cosmic string network. These analysis are thus focused on the definition of the differential equations defining the network and on the numerical evolution of these equations.
	This class of works is the most diffused in literature. See for example~\cite{AvgoustidisShellard2008,AvgoustidisShellard2009,TyeWassermanWyman2005,CuiMartinMorrisseyEtAl2008}.
\end{itemize}
Of course every class of simulations presents some merits and flaws. Field-oriented simulations are easy to be implemented using lattice techniques but they may require an excessive computational power and they don't allow easily the introduction of a mechanism to treat GW emissions from the string network. String-oriented simulations give a direct control on the evolution of the strings, and as they are expressed in terms of Nambu action, they admit the possibility of introducing a coupling between strings and gravity. At the same time these simulations are difficult to be set and they may be time consuming. Finally phenomenogical simulations are faster and less numerically demanding than the other classes. On the other hand they do not give a direct control on the evolution of the network.

In this paper we present a simple model, extremely similar to the one treated in~\cite{Spergel,CopelandSaffin2005}. Actually we aim at defining a model that can be easily studied by means of lattice techniques without requiring an excessive computational power. In particular we focus on the study of two problems:
\begin{itemize}
	\item The consequences of introducing a prescription to deal with a non trivial issue of the models discussed in~\cite{Spergel,CopelandSaffin2005}. 
	\item The stabilitity of the scaling solution when we consider models with a large number of string generators.
\end{itemize}
To have a better understanding of the evolution of the lattice we also introduce some new observable quantities to grasp some more refined informations on the behavior of the network. We stress that this paper should be intended as a first step towards the possibility of producing large scale simulations on realistic non-Abelian models. In the case of our interest a scaling solution appears to be reached.

The paper is structured as follows: in Section~\ref{sec:The-model-and} we present the models to generate the network of cosmic strings and we discuss its discretization on a lattice. In 
Section~\ref{sec:Numerical-results} we present the results of our simulations 
and in Section~\ref{sec:Conclusions} we draw our conclusions.

\section{\label{sec:The-model-and}The model and its dynamics}

In this paper we consider a slight variation of the model discussed by Spergel 
and Pen in~\cite{Spergel}. To produce non-Abelian cosmic strings the 
vacuum manifold for the theory should have a non-Abelian fundamental group. To 
construct an example of a space with this property, we start by considering the 
space defined as $\bar{\mathcal{M}}_{N}=[0,\pi]\times I_{N}$ where $I_{N}$ 
denotes the set $\{0,1,\dots,N-1\}$. Let us define  the equivalence relationship $*$
on $\bar{\mathcal{M}}_{N}$, such that $\forall\ i,i^{\prime}\in I_{N}$, 
$(0,i)*(0,i^{\prime})$ and $(\pi,i)*(\pi,i^{\prime})$. Finally let us consider 
the space $\mathcal{M}_{N}$ defined as 
$\mathcal{M}_{N}\equiv\bar{\mathcal{M}}_{N}/*$. This space is homotopic to the bouquet of $N-1$ circles whose fundamental group is $F_{N-1}$, where $F_{N}$ is the free group of rank $N$ 
which is a non-Abelian group. As the fundamental group of $\mathcal{M}_{N}$ is non abelian, by fixing $\mathcal{M}_{N}$ to be the vacuum set for our theory, the model will lead to the production of non-Abelian cosmic strings. It is thus sufficient to define scalar field theory with the Higgs field $\phi$ taking values in $\mathcal{M}_{N}$.
It is important to point out that for $N > 2 $, $\mathcal{M}_{N}$ is not a manifold as it does not exist $k \in \mathbb{N}$ such that $\mathcal{M}_{N}$ is diffeomorphic to $\mathbb{R}^{k}$ in $0$ and $\pi$. As discussed in this section, we will ignore this issue by fixing a rule that properly defines the dynamics in these two points.

Let us consider the theory defined by the action:
\begin{equation}
\label{eq:scalar_theory}
S=\frac{1}{2}\int \mathrm{d}^4 x \sqrt{-g} g^{\mu \nu} \partial_\mu \phi \partial_\nu \phi,
\end{equation}
with $\phi \in \mathcal{M}_{N}$ and with the space-time interval defined as:
\begin{equation}
\label{interval}
ds^{2}=g_{\mu\nu}dx^{\mu}dx^{\nu}=g_{00}dt^{2}-a^{2}(t)\,d\vec{x}\cdot d\vec{x}.
\end{equation}
Notice that for $g_{00}=1$, $t$ in Equation~\eqref{interval} is identified with 
the cosmic time while for $g_{00}=a^{2}(t)$ it corresponds to the conformal time. For the scope of this work we need to define a spatial discretization of the 
theory on a lattice. To implement this procedure, let us denote with $\phi_{\vec{x},t}\in{\cal M}_{N}$ the value of the fild $\phi(\vec{x},t)$ in the site with position $\vec{x}$ of a cubic lattice 
with coordinate spacing $h$ at the time $t$. Let us define the parameterization 
$\phi_{\vec{x},t}=\left(\theta_{\vec{x},t},i_{\vec{x},t}\right)$, where the integer 
number $i_{\vec{x},t} \in \{0,1,\dots,N-1\}$ is used to denote one of branches in $\mathcal{M}_{N}$ and the 
real variable $\theta \in [0,\pi]$ is used to identify a point in the 
corresponding branch. Let us define:
\begin{equation}
\Delta\left(\phi_{\vec{x},t},\phi_{\vec{y},t}\right)  \equiv \left\{ \begin{array}{ccc}
e^{i\theta_{\vec{x},t}}-e^{i\theta_{\vec{y},t}} & \qquad \mbox{if} & \qquad  i_{\vec{x},t}=i_{\vec{y},t}\\
e^{-i\theta_{\vec{x},t}}-e^{i\theta_{\vec{y},t}} & \qquad  \mbox{if} & \qquad  i_{\vec{x},t}\neq i_{\vec{y},t}
\end{array}\right.
\end{equation}
then the action can be expressed as:
\begin{equation}
\label{discrete_action}
S=\frac{1}{2}\int dt\sum_{\vec{x},i}h^{3}\sqrt{-g}\left[g^{00}\dot{\theta}_{\vec{x},t}^{2}+h^{-2}g^{ii}(t)\Delta\left(\phi_{\vec{x}+\vec{e}_{i},t},\phi_{\vec{x},t}\right)\Delta\left(\phi_{\vec{x}+\vec{e}_{i},t},\phi_{\vec{x},t}\right)^{*}\right],
\end{equation}
where $\vec{x}+\vec{e}_{i}$ is used to denote the closest site to $\vec{x}$ in the direction $\vec{e}_{i}$. Defining $\sigma_{\vec{x},\vec{y},t}$ as:
\begin{equation}
\sigma_{\vec{x},\vec{y},t} \equiv \left\{ \begin{array}{ccc}
1 & \qquad \mbox{if} & \qquad  i_{\vec{x},t}=i_{\vec{y},t}\\
-1 & \qquad  \mbox{if} & \qquad  i_{\vec{x},t}\neq i_{\vec{y},t}
\end{array}\right. \ ,
\end{equation}
 the lagrangian density simply reads:
\begin{equation}
\label{eq:LAGRANGIAN}
{\cal L}=\frac{1}{2}h^{3}a^{3}\sum_{\vec{x}}\left[\frac{1}{\sqrt{g_{00}(t)}}\dot{\theta}_{\vec{x},t}^{2}+\sqrt{g_{00}(t)}\frac{2}{a^{2}h^{2}}\sum_{i}\cos\left(\theta_{\vec{x},t}-\sigma_{\vec{x} , \vec{x}+\vec{e}_{i},t}\,\theta_{\vec{x}+\vec{e}_{i},t}\right)\right].
\end{equation}
We can thus compute the Hamiltonian density:
\begin{equation}
\label{eq:HAMILTONIAN}
H=\sqrt{g_{00}(t)}\sum_{\vec{x}}\left[\frac{\pi_{\vec{x},t}^{2}}{2h^{3}a^{3}}-ah\sum_{i}\cos\left(\theta_{\vec{x},t}-\sigma_{\vec{x},\vec{x}+\vec{e}_{i},t}\,\theta_{\vec{x}+\vec{e}_{i},t}\right)\right].
\end{equation}
Finally we obtain the equations of motion:
\begin{eqnarray}
\frac{1}{\sqrt{g_{00}(t)}}\dot{\theta}_{\vec{x},t} & = & \frac{1}{h^{3}a^{3}}\pi_{\vec{x},t}, \label{eq:Hq}\\
\frac{1}{\sqrt{g_{00}(t)}}\dot{\pi}_{\vec{x},t} & = & -ah\sum_{\vec{y}}\sin\left(\theta_{\vec{x},t}-\sigma_{\vec{x},\vec{y}}\,\theta_{\vec{y},t}\right),\label{eq:Hp}
\end{eqnarray}
where we have defined $\vec{y}_i = 
\vec{x}+\vec{e}_{i}$ in Eq.\eqref{eq:Hp} to simplify the notation. Notice that Eq.\eqref{eq:Hq} and Eq.\eqref{eq:Hp} are not 
giving any information about the evolution of the variables $i_{\vec{x},t}$. This 
issue is connected to the fact that $\mathcal{M}_{N}$ is not a manifold for 
$N>2$, which is the non-Abelian case. In particular, it is not be possible to 
properly choose well defined coordinates in the neighborhood of $\theta=0$, 
$\theta=\pi$. To solve this issue, we have to fix a prescription to assign a new 
value to the discrete index $i_{\vec{x},t}$, when $\theta$ assumes one of the two 
critical values $\theta_{\vec{x},t}=0$, $\theta_{\vec{x},t}=\pi$. The procedure followed 
in~\cite{Spergel} consists in fixing an ordering to separately consider 
the evolution induced by every term in the sum on the right side of 
Eq.~\eqref{eq:Hp}. With this prescription the evolution is decomposed into a 
series of elementary steps with $N = 2$. As in this case $\mathcal{M}_{N}$ 
is a manifold, the evolution is completely specified. It is however important 
to stress that this procedure is order dependent. As, a priori, no preferred 
order can be chosen, this procedure is arbitrary and can not be conceived as 
completely general. To avoid this arbitrariness, in our work we fix a different 
prescription to solve the problem for $N>2$. 

The numerical evolution of the 
system is performed by means of a symplectic integrator~\cite{LEAPFROG} for a 
time dependent Hamiltonian~\cite{HTDEPENDENT}. This method is correct up to 
second order in the time step $\Delta t$. Whenever a time step in the evolution 
produces $\theta_{\vec{x},t+\Delta t}<0$ or $\pi < \theta_{\vec{x},t+\Delta t}$, 
we perform the replacement $\theta_{\vec{x},t+\Delta 
	t}\equiv-\theta_{\vec{x},t+\Delta t}$ or $\theta_{\vec{x},t+\Delta 
	t}\equiv2\pi-\theta_{\vec{x},t+\Delta t}$ respectively. With this prescription 
we secure $\theta_{\vec{x},t+\Delta t}$ to be in its existence domain. In 
addition to this procedure we have to define a rule to deal with the possible 
change in discrete index $i_{\vec{x},t}$. To fix the proper value for 
$i_{\vec{x},t+\Delta t}$ we perform a comparison between the potential terms of 
Eq.~\eqref{eq:HAMILTONIAN}. We sum up the contributions due to the same discrete 
index and we select the $\bar{i}$ associated to the greatest one. At this point 
it seems reasonable to assign the value $\bar{i}$ to the discrete variable $i_{\vec{x},t+\Delta t}$.

\subsection{\label{sub:The-string-density}The string density}

To compute the string density during the evolution we have to define a procedure to measure the number of plaquettes in the lattice that are pierced by a string. Let us consider the square with vertices $\vec{x}_{1}=\vec{x}$, $\vec{x}_{2}=\vec{x}+\hat{e}_{i}$,
$\vec{x}_{3}=\vec{x}+\hat{e}_{i}+\hat{e}_{j}$, $\vec{x}_{4}=\vec{x}+\hat{e}_{j}$, with $i\neq j$. In order to evaluate the ``flux'' associated with the plaquette we introduce:
\begin{eqnarray}
q_{ij} & = & \theta_{j}-\theta_{i},\\
\overline{q}_{ij} & = & \pi\Theta\left(\theta_{i}+\theta_{j}-\pi\right)-\theta_{i},
\end{eqnarray}
where the discrete indices $i,j = \{ 1,2,3,4\}$ are used to label the 4 vertices of the plaquette, $\theta_{i} \equiv \theta_{\vec{x}_i,t}$ denotes the value of $\theta_{\vec{x},t}$ in the site $i$ and $\Theta$ is the Heaviside step function. These two quantities should be discussed in relation with the branch configuration assumed by the field. In fact we should face several possibilities and it is better to consider each case separately. As this procedure is defined to compare values of the field on different sites at fixed time $t$, for the rest of this section we set $i_{\vec{x},t} = i_{\vec{x}}$ to lower the notation. 

\begin{figure}[htp]
	\begin{centering}
		\includegraphics[width=0.8\columnwidth]{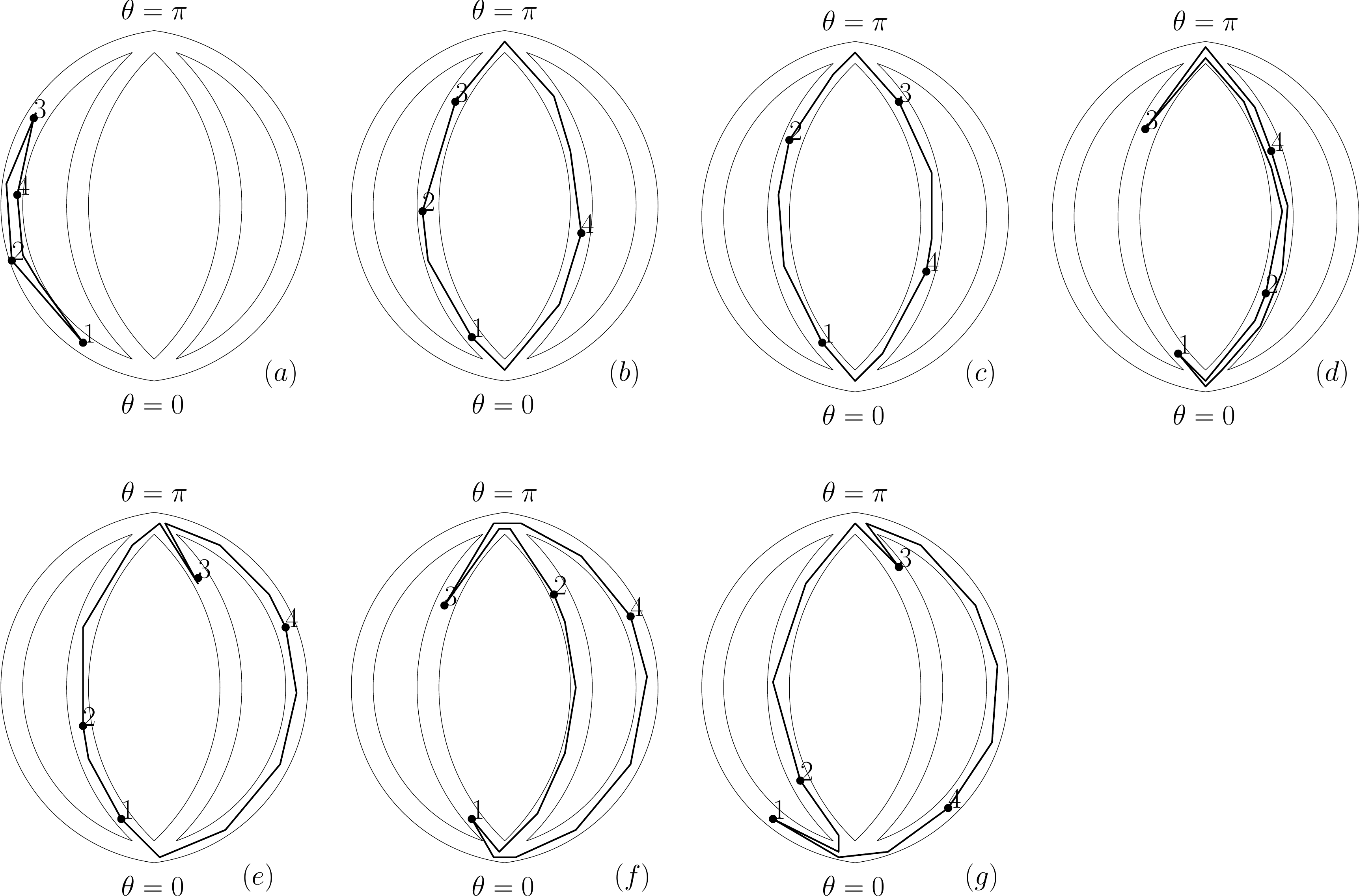}
	\end{centering}
	\caption{\label{fig:branches} We present the set of all possible configurations assumed by the field on the vertices of a plaquette. The numbered points are used to indicate the value of the fields on the four nodes. Notice that two nodes are always connected following
		the shorthest path. Non-contractible loops are associated with the  piercing of a string through the plaquette. Non-contractible loops can be originated in $(b)$, $(c)$, $(e)$, $(f)$ $(g)$.}
\end{figure}

In the simplest case we have 
$i_{\vec{x}_{1}}=i_{\vec{x}_{2}}=i_{\vec{x}_{3}}=i_{\vec{x}_{4}}$ and thus the 
four fields are defined on the same branch. The corresponding path in the vacuum 
space $\mathcal{M}_{N}$ is expected to be homeomorphic to the configuration 
represented in in Fig~\ref{fig:branches}-$(a)$. In this case, the expression for 
the total displacement is simply given by $q_{12}+q_{23}+q_{34}+q_{41}$, which 
is obviously equal to zero. This situation is thus trivial and does not need any 
further characterization as we can directly conclude that strings are not 
allowed to pierce the plaquette. Let us consider the case 
$i_{\vec{x}_{1}}=i_{\vec{x}_{2}}=i_{\vec{x}_{3}}\neq i_{\vec{x}_{4}}$. This 
situation is depicted in Figure~\ref{fig:branches}-$(b)$. In this situation it 
is useful to consider the quantity 
$\imath_Q \equiv q_{12}+q_{23}+\overline{q}_{34}-\overline{q}_{14}$. If the 
first branch is completely traversed in the positive direction $\imath_Q$ 
assumes the value $\pi$. On the contrary, if the first branch is traversed in 
the negative direction it is equal to $-\pi$. In any other configuration 
$\imath_Q=0$. A quantity with the
same properties, $\overline{q}_{41}-\overline{q}_{43}$, can be defined on the second branch. Notice that every string can be associated with a ``color'' and an ``anti-color''. The color is determined by the index of the branch traversed in the positive direction and the anticolor is given by the index of the branch traversed in the negative direction. In this section, to be consistent with the examples shown in Figure~\ref{fig:branches}, the color of the string is determined by the branch of the first vertex i.e. $i_{\vec{x}_{1}}$ and the anti-color is determined by the branch of the fourth vertex $\overline{i}_{\vec{x}_{4}}$. In terms of these quatities the probability for a string of type $i_{\vec{x}_{1}}\overline{i}_{\vec{x}_{4}}$ to pierce the plaquette is given by:
\begin{equation*}
	{\cal J}_{b}=\int\prod_{i=1}^{4}d\theta_{i}\,P\left(\theta_{1},\theta_{2},\theta_{3},\theta_{4}\right)\Theta\left(q_{12}+q_{23}+\overline{q}_{34}-\overline{q}_{14}\right)\Theta\left(\overline{q}_{43}-\overline{q}_{41}\right),
\end{equation*}
where $P\left(\theta_{1},\theta_{2},\theta_{3},\theta_{4}\right)$ is the probability distribution for $\theta_{1},\theta_{2},\theta_{3},\theta_{4}$. If we are not interested in distinguish between $i_{\vec{x}_{1}}\overline{i}_{\vec{x}_{4}}$ and $i_{\vec{x}_{4}}\overline{i}_{\vec{x}_{1}}$ the probability for a string to pierce the plaquette in the configuration of Figure~\ref{fig:branches}-$(b)$ is given by:
\begin{eqnarray*}
	{\cal I}_{b} & = & \frac{1}{\pi^{2}}\int\prod_{i=1}^{4}d\theta_{i}\,P\left(\theta_{1},\theta_{2},\theta_{3},\theta_{4}\right)\left(q_{12}+q_{23}+\overline{q}_{34}-\overline{q}_{14}\right)\left(\overline{q}_{43}-\overline{q}_{41}\right),\\
	{\cal I}_{b} & = & \left\langle \left(g_{34}-g_{14}\right)^{2}\right\rangle,
\end{eqnarray*}
where we defined $g_{ij}\equiv \Theta\left(\theta_{i}+\theta_{j}-\pi\right)$ and $\left\langle X\right\rangle $ is used to denote the expectation value of $X$. It is possible to show that for the other generic configurations we can obtain similar expressions for ${\cal I}$. In particular for the configurations of Figure~\ref{fig:branches} the results can be expressed as expectation values of $g_{ij}$ as:
\begin{eqnarray*}
	{\cal I}_{c} & = & \left\langle \left(g_{14}-g_{23}\right)^{2}\right\rangle , \\
	{\cal I}_{d} & = & \left\langle \left(g_{12}-g_{23}-g_{14}+g_{34}\right)^{2}\right\rangle ,\\
	{\cal I}_{e} & = & \left\langle g_{14}^{2}+g_{23}^{2}+g_{34}^{2}-g_{23}g_{34}-g_{14}g_{23}-g_{14}g_{34}\right\rangle ,\\
	{\cal I}_{f} & = & \left\langle 
	g_{12}^{2}+g_{34}^{2}+g_{14}^{2}+g_{23}^{2}+g_{14}g_{23}-2g_{14}g_{34}-g_{23}g_{
		34}-g_{12}g_{14} \right. \nonumber \\
	& & - \left. 2g_{12}g_{23}+g_{12}g_{34}\right\rangle 
	,\\
	{\cal I}_{g} & = & \left\langle g_{12}^{2}+g_{14}^{2}+g_{23}^{2}+g_{34}^{2}-g_{12}g_{14}-g_{12}g_{23}-g_{14}g_{34}-g_{23}g_{34}\right\rangle .
\end{eqnarray*}
It is interesting to notice that by using the isotropy of the lattice we can proceed with a further simplification of these expressions: 
\begin{eqnarray*}
	{\cal I}_{b} & = & 1-2\Gamma_{2} ,\\
	{\cal I}_{c} & = & 1-2\Gamma_{1} ,\\
	{\cal I}_{d} & = & 2-8\Gamma_{2}+4\Gamma_{1} ,\\
	{\cal I}_{e} & = & \frac{3}{2}-2\Gamma_{2}-\Gamma_{1} ,\\
	{\cal I}_{f} & = & 2+2\Gamma_{1}-6\Gamma_{2} ,\\
	{\cal I}_{g} & = & 2-4\Gamma_{2} ,
\end{eqnarray*}
where $\Gamma_{1}=\left\langle g_{12}g_{34}\right\rangle $, $\Gamma_{2}=\left\langle g_{12}g_{14}\right\rangle $. Notice that to derive these expressions we used $\left\langle g_{ij}^{2}\right\rangle =\left\langle g_{ij}\right\rangle =1/2$ that has been deduced using the symmetry of the theory under the transformation $\theta_{i}\rightarrow\pi-\theta_{i}$. Finally we define the string density as the number of strings per lattice volume unit. It is clear that this quantity is equal to:
\begin{equation}
\rho=3\sum_{k\in\left\{ b,c,d,e,f,g\right\} }{\cal P}_{k}{\cal I}_{k}\label{eq:rhoanalytic}
\end{equation}
where ${\cal P}_{k}$ is the probability of each configuration in
Figure~\ref{fig:branches} (summed over all the possible values of nodes' 
fields). As the evolution of our system is completely deterministic, the only 
stochastic factor
is given by the initial configuration.

\section{\label{sec:Numerical-results}Numerical results}

The simulations are implemented on $300^{3}$, $600^{3}$ and $800^{3}$
spatial lattices with periodic boundary conditions. This analysis
is realized in a radiation dominated epoch, which is consistent with the 
assumption that the string energy density never happens to contribute in a 
significant way to the evolution of the scale factor described by Friedmann 
equation.

We consider an initial proper lattice spacing $d=ah$ greater than the 
correlation length $\xi$ associated to the field. The latter is supposed to be 
of the same order of the Hubble radius $R_{H}$. Under these 
assumptions we can give random initial values for the field $\theta_{i}$ to the 
lattice sites. We choose always an initial configuration
with zero momenta $\pi_{\vec{x},t}$: in principle it will be interesting to extend the study by including several different initial conditions, as in~\cite{CopelandSaffin2005}. However this will be quite time consuming and our focus is mainly to obtain a fast benchmark. 

Notice that in a radiation dominated epoch, the physical separation between sites increases proportionally to $t^{1/2}$ while the Hubble radius $R_{H}$ evolves accordingly with $R_{H}\sim t$. The choice $\xi\sim R_{H}$ then provides a natural way to fix an ending to our numerical simulation: when the horizon is of the same order of the lattice itself, $R_{H}\simeq N_{s}ah$, boundary conditions can strongly affect the evolution and thus we are forced to conclude the simulation. 

\begin{figure}[htp]
	\begin{centering}
		\includegraphics[width=0.8\columnwidth]{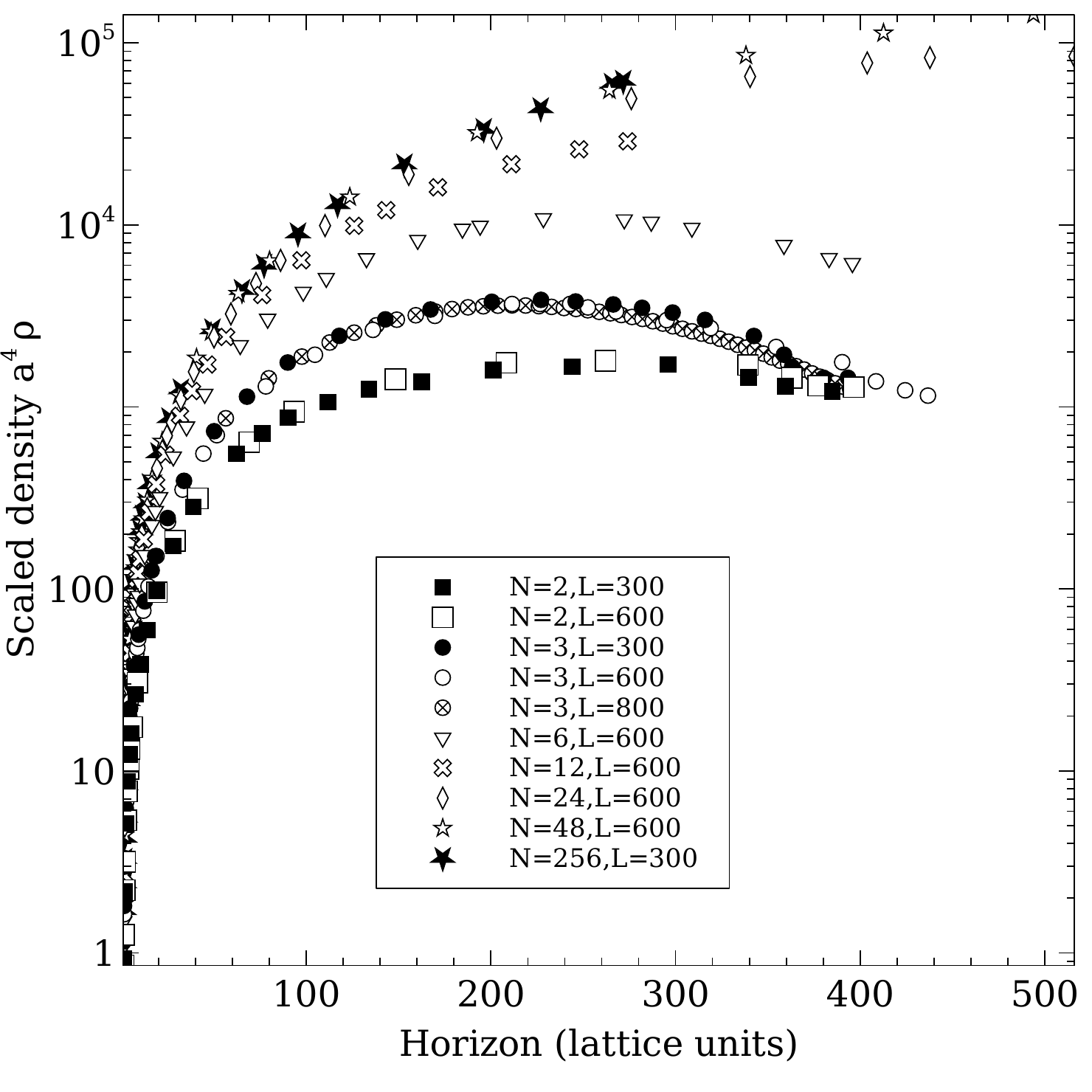}
	\end{centering}
	\caption{\label{fig:lattice}Results of the numerical simulations respectively
		in linear and logarithmic plot on a $300\times300\times300$ lattice.}
\end{figure}

Figure~\ref{fig:lattice} shows the outcome of numerical simulations.
The measured string density multiplied by the scale factor $a^{4}$
is plotted as a function of the horizon, given in lattice units. Notice that 
because of the introduction of the $a^{4}$ factor, quantities that scale 
similarly to the radiation energy density should appear as constants. The 
simulations in the physically significant region show different behaviors for 
networks associated with different values for $N$. As expected, the Abelian case 
$N=2$ (values represented by squares) rapidly reaches the scaling behavior and 
thus the string density can not dominate over radiation energy density. The same 
result appears to be true for the $N=3$ non-Abelian case (values represented by 
circles). It is important to stress that in this case the scaling behaviour is 
reached for bigger values of the horizon in lattice units. The $N=6$ and $N=12$ 
cases are quite similar.

\begin{figure}[htp]
	\begin{centering}
		\includegraphics[width=0.8\columnwidth]{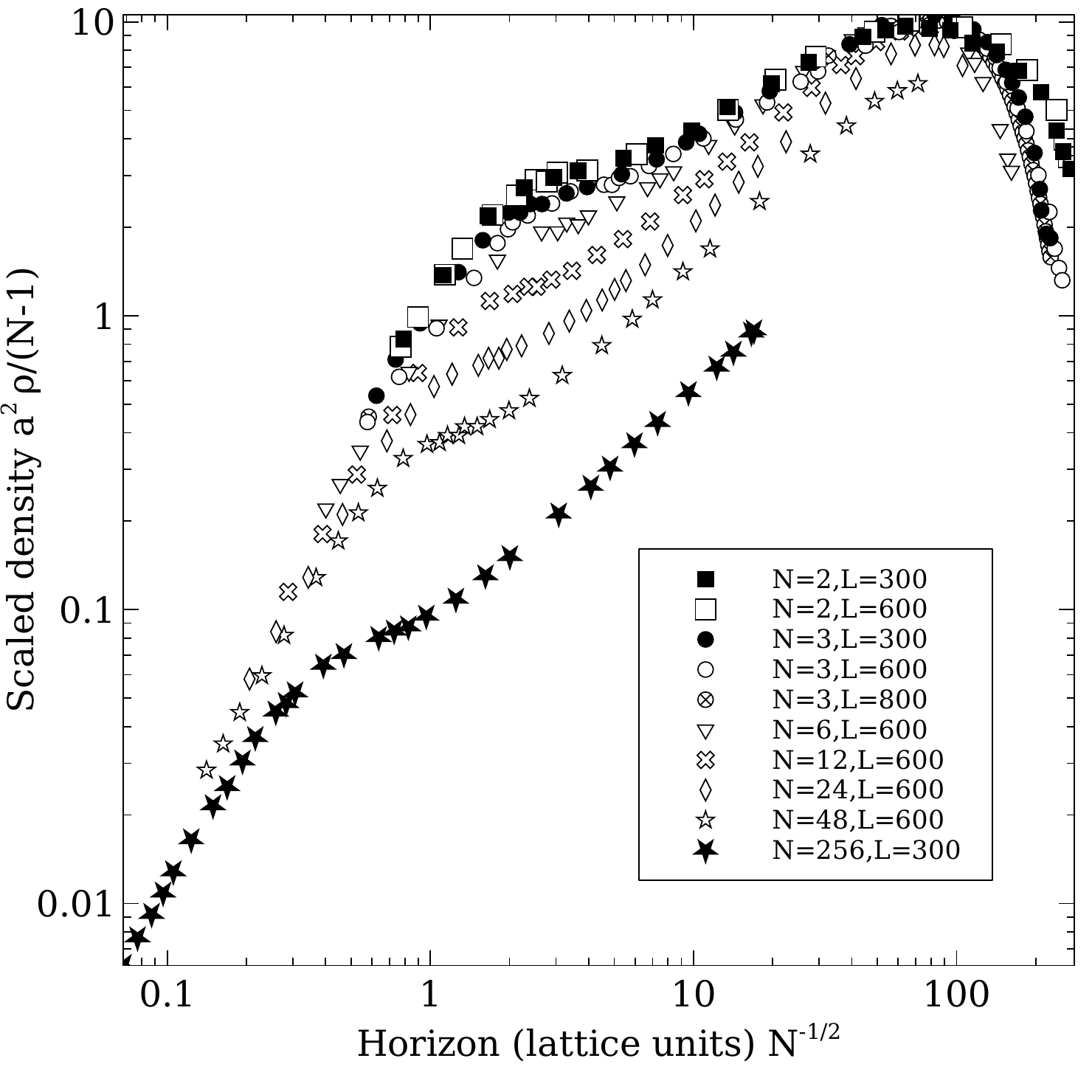}
	\end{centering}
	\caption{\label{fig:universality}A rescaled plot of the results show in Figure~\ref{fig:lattice}
		respectively in linear and logarithmic scale. Interestingly the maxima
		of the curves are thickening in the same region.}
\end{figure}

For the purposes of our work we are also interested in studying models with bigger values $N$. In these cases it is not possible to appreciate the occurrence of a scaling solution. However, given the behavior of the system for the cases with small values for $N$, it seems natural to suppose that a scaling regime will be reached for big values of the horizon in lattice units. In particular, these values will be outside of the observable range for the lattice used to implement the situation. To support this hypotesis it is then interesting to observe the graphs of Figure~\ref{fig:universality}. In these plots, we plotted the scaled string density $a^{2}\rho\left(N-1\right)^{-1}$ versus the scaled horizon size in lattice units $\left(H/d\right)N^{-1/2}$. In a maximum of this plot, the string energy density is approximatively scaling as $a^{-2}$. As the rescaled positions of the maxima accumulate around the same value for the horizon it seems natural to assume that this condition is reached after a time interval that is proportional to $N^{1/2}$.

From Figure~\ref{fig:universality} it is possible to distinguish three different
regimes during the evolution of the string density. In a short inital phase the string density $\rho$ is approximately constant. This is a consequence of choosing an initial conditions with zero momenta. During this epoch kinetic energy still has to be transferred to the strings and thus they are not able to move. In this phase string can not interconnect\footnote{In
	a non-Abelian model the interconnection of two strings can originate
	a new ``bridge''} and thus there is no mechanism to chop off long strings into 
shorter loops. The length of this phase should be approximately $N$-independent. 
However, to be more accurate, we can stress that the initial probability for two 
sites to be in different branches is proportional to $1-N^{-1}$. This implies 
that initial potential energy available should sightly increase with $N$. This 
appears to be in agreement with results shown in Figure~\ref{fig:universality}. 
After the initial phase, strings starts to move and their density start to 
decrease, until the maximum is reached for at a time $t^{*}$ proportional to 
$N^{1/2}$. We still have not a rigorous theoretical argument to explain this 
relationship. In the latter phase the scaled string density finally starts to 
fall. 

The initial strings densities vary accordingly to the number of generators. We can evaluate the initial energy density in lattice volume units by using the expressions derived in Sec.\ref{sub:The-string-density}. If the initial configuration is completely random, the probability distribution for $\theta_{i}$
on a plaquette is simply given by $P\left(\theta_{1},\theta_{2},\theta_{3},\theta_{4}\right)=\pi^{-4}$
and the correlations $\Gamma_{i}$ can easily be evaluated, obtaining
\begin{eqnarray*}
	\Gamma_{1}^{(0)} & = & \frac{1}{\pi^{4}}\int\prod_{i=1}^{4}d\theta_{i}\Theta\left(\theta_{1}+\theta_{2}-\pi\right)\Theta\left(\theta_{3}+\theta_{4}-\pi\right)=\frac{1}{4}\\
	\Gamma_{2}^{(0)} & = & \frac{1}{\pi^{4}}\int\prod_{i=1}^{4}d\theta_{i}\Theta\left(\theta_{1}+\theta_{2}-\pi\right)\Theta\left(\theta_{1}+\theta_{4}-\pi\right)=\frac{1}{3}
\end{eqnarray*}
while the configuration probabilities ${\cal P}_{k}^{(0)}$ are given
in Table~\ref{tab:confprob}. Using Equation~\eqref{eq:rhoanalytic}
we find:
\begin{equation}
\rho=3\sum_{k\in\left\{ b,c,d,e,f,g\right\} }{\cal P}_{k}^{(0)}{\cal I}_{k}^{(0)}=2\left(1-\frac{1}{N}\right),
\end{equation}
which is in good agreement with the simulation results. We note that
the string density in the $a^{-2}$ regime seems to be approximately
independent from $N$. In fact we have $a^{2}(t^{*})\propto t^{*}\propto N$:
if we multiply this factor by the initial string density we obtain
the normalization factor used in Figure~\ref{fig:universality},
where all the maxima have more or less the same value.

\begin{table}
	\begin{centering}
		\begin{tabular}{c|ccccccc}
			& $a$ & $b$ & $c$ & $d$ & $e$ & $f$ & $g$\\ 
			\hline
			${\cal P}_{k}^{(0)}$ & $\frac{1}{N^{4}}$ & $\frac{4N(N-1)}{N^{4}}$ & $\frac{2N(N-1)}{N^{4}}$ & $\frac{N(N-1)}{N^{4}}$ & $\frac{4N(N-1)(N-2)}{N^{4}}$ & $\frac{2N(N-1)(N-2)}{N^{4}}$ & $\frac{N(N-1)(N-2)(N-3)}{N^{4}}$\\ [1.2ex]
			${\cal I}_{k}^{(0)}$ & $0$ & $\frac{1}{3}$ & $\frac{1}{2}$ & $\frac{1}{3}$ & $\frac{7}{12}$ & $\frac{1}{2}$ & $\frac{2}{3}$\\[1.2ex]
		\end{tabular}
	\end{centering}	
	\caption{\label{tab:confprob}The probabilities relevant for the calculation
		of the string density for each configuration in Figure~\ref{fig:branches}
		in a completely random configuration.}
\end{table}

We extract from the simulations other observables to get more information
on the evolution of the networks. As explained in Sec.\ref{sub:The-string-density}, once we have fixed a prescription
to move along the elementary plaquette, the model allows to associate
a string with a ``color'' and an ``anti-color''. The number of different strings for a $N$ branches model is $N(N-1)$. If $\rho_{c\bar{d}}$ be the density of strings associated with a certain color/anti-color pair, we define the asymmetry: 

\begin{equation}
{\cal A}\equiv\sum_{c\neq d}\frac{\left|\rho_{c\bar{d}}-\rho_{d\bar{c}}\right|}{\rho_{c\bar{d}}+\rho_{d\bar{c}}}\label{eq:ASYMMETRY}.
\end{equation}

\begin{figure}[htp]
	\includegraphics[width=0.5\columnwidth]{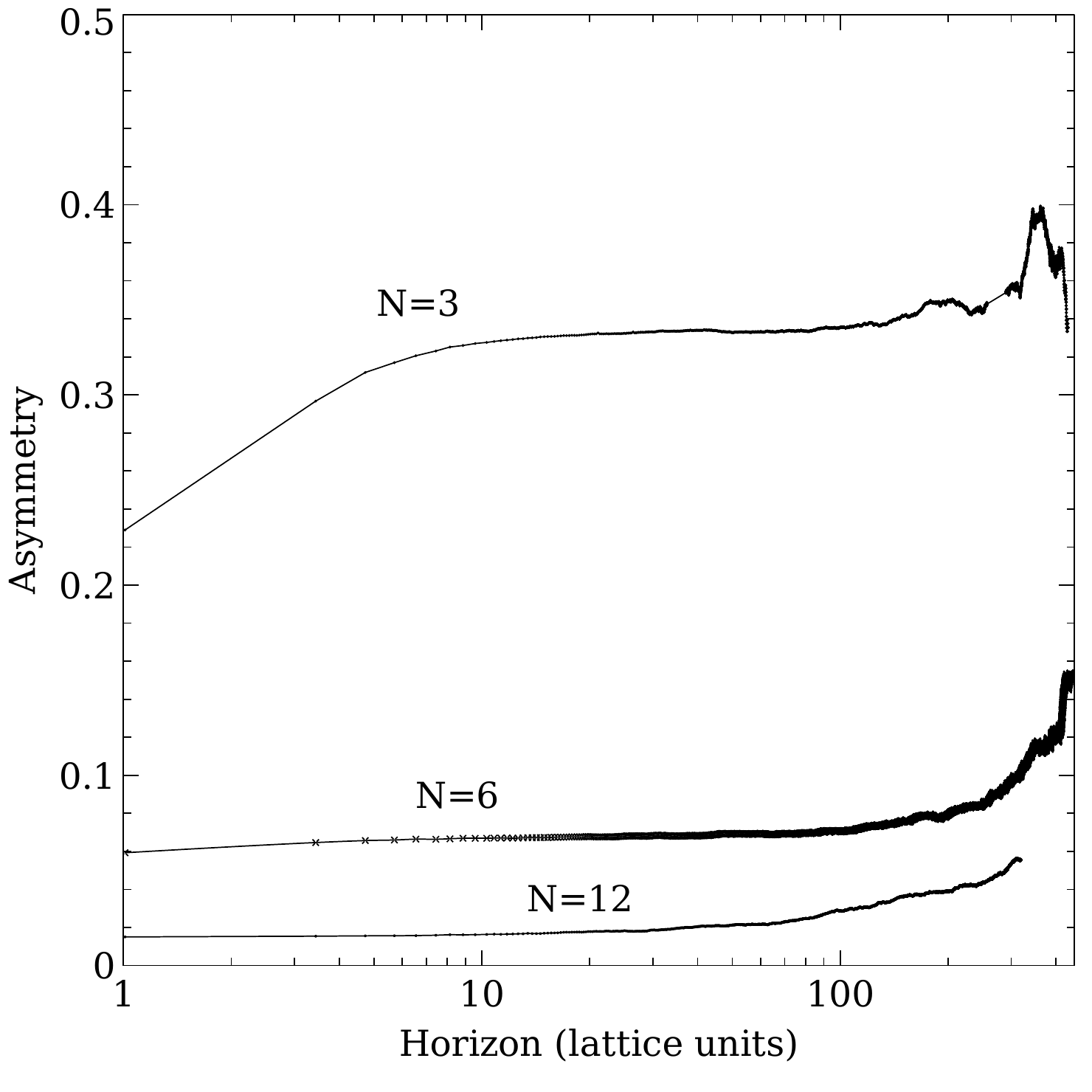}
	\includegraphics[width=0.5\columnwidth]{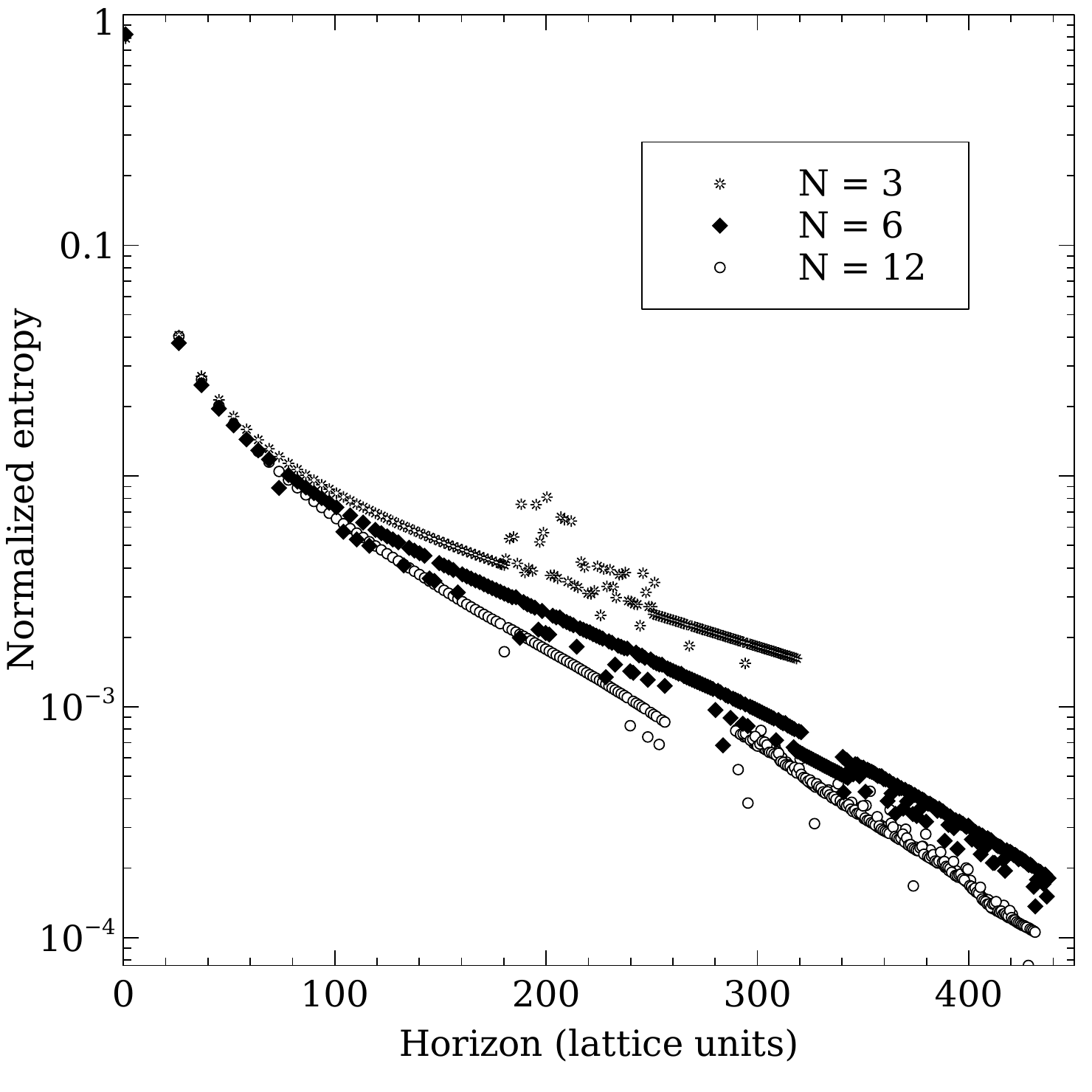}
	\caption{\label{fig:asymmetry_entropy}The left plot shows the evolution, as a function of the horizon expressed in lattice units, of the asymmetry parameter defined by Eq.~\eqref{eq:ASYMMETRY} for $N=3,6,12$ on a $600^{3}$ lattice. The right plot shows the same evolution for the normalized entropy defined in Eq.~\eqref{eq:ENTROPY}.}
\end{figure}

Notice that loops that are fully contained into the lattice does not contribute
to the numerator of this quantity. On the other hand ${\cal A}$ quantifies all 
of those structures produced by non-Abelian interactions such as Zipper, 
Bridges~\cite{McGraw:1996py} and loops encircling the whole lattice. As it is 
possible to see from the graph on the left of Figure~\ref{fig:asymmetry_entropy} 
the asymmetry increases when the string density falls off. Such an effect is 
clearly due to the particular conditions required to produce strings' 
annihilation. As a matter of fact loops
can collapse in a finite time because of their dynamics but on the
other hand such a mechanism does not affect non-Abelian structures
that require more refined circumstances to disappear. It would be interesting to compare these result with the study of Bettencourt and Kibble~\cite{BettencourtKibble1994} who have analyzed the conditions to create zippers for networks of type I strings.

Let us define the total string density $\rho_{TOT}\equiv\sum_{c\neq d}\rho_{c\bar{d}}$, and the ``entropy'' associated with a certain string configuration: 

\begin{equation}
{\cal S}\equiv-\sum_{c\neq d}\frac{\rho_{c\bar{d}}}{\rho_{TOT}}\ln\frac{\rho_{c\bar{d}}}{\rho_{TOT}}\label{eq:ENTROPY}.
\end{equation}

In the initial and completely random configuration, ${\cal S}$ has the largest value, which corresponds to a configuration where strings
are equally divided over all the possible color/anti-color couples.
A plot of $S$ is shown on the right in Figure~\ref{fig:asymmetry_entropy}. As expected, the random initial conditions induce a high value for ${\cal S}$. As string dynamics tend to rearrange the network, we
depart from the initial random configuration and this causes ${\cal S}$ to fall down. In the last part of the evolution we notice some fluctuations on ${\cal S}$. This effect is due to the decreasing of $\rho_{TOT}$. Such a process indeed implies that small variations within the lattice can produce relevant effects on ${\cal S}$.

\section{\label{sec:Conclusions}Conclusions}

The results obtained in this paper are basically in agreement with the existing literature. Most of the existing field-oriented~\cite{Spergel,CopelandSaffin2005,HindmarshSaffin2006}, string-oriented~\cite{McGraw1,McGraw:1996py} and phenomenological simulations~\cite{AvgoustidisShellard2008,AvgoustidisShellard2009,TyeWassermanWyman2005,CuiMartinMorrisseyEtAl2008} agree on the appearence of a scaling solution.  

We obtained some interesting results on the behavior of the system as we variate $N$, number of strings' generators, in particular promising conclusions are drawn for the case of large values of $N$. At a first glance numerical results in a radiation dominated epoch seem to produce different outcomes for different values of $N$. The Abelian case, as expected, appears to scale in agreement with the hypothesis of a string energy density $\propto a^{-4}$, that never happens to dominate Friedmann equation. The case of an heavily non-Abelian network evolves differently. Its energy density does not seem to scale, which will be inconsistent with the hypothesis of a radiation dominated universe. 

A more refined analysis of the results obtained for different values of $N$ shows some hints of an underlying structure. In particular this can be appreciated by looking at the plots shown in Figure~\ref{fig:universality}. These plots show that the initial phase, where string energy density increases appears to last for a period proportional to $N^{1/2}$. After the maximum is reached, the energy density decreases and approaches the expected scaling solution. Moreover, it is interesting to notice that the curves describing the energy densities associated with networks for different values of $N$, appear to be quite similar after a rescaling with a certain power of the number of generators. The good accordance shown in the plots of Figure~\ref{fig:universality} furnishes a scheme to produce a rough estimate of the value of the horizon in lattice units corresponding to the occurrence of the inflection point. For a model with $N=256$ we may guess a shift of order $16$. To avoid volume effects, we should then produce simulations on a $\sim4000$ times greater lattice. Obvious computational problems arise.

Figure~\ref{fig:lattice} can be compared with Figure~5 in~\cite{CopelandSaffin2005}, which deals with a similar model. In both cases we see evidence of scaling and of a string energy density which increases with $N$ (roughly in a linear way in our case).

In~\cite{CuiMartinMorrisseyEtAl2008} a different model with several string species parameterized by an integer $N_{max}$ is studied, for $N_{max}$ values up to $200$. Scaling is observed, and the time needed to reach the scaling regime is found to be described by $t\propto N_{max}^\alpha$, however with a value for $\alpha$ much larger than the $\alpha=1/2$ one we observed. The string energy density is an increasing function of $N_{max}$, though only a logarithmic one.

It is fair to point out that in the simulations presented in this work we assumed a radiation dominated universe, and thus we have never taken into account the contribution of the string energy density into the evolution of the scale parameter. While this effect does not appear to produce significant deviations in the case of small values for $N$, it can be interesting to understand its consequences in the case of $1 \ll N$. This more general case can be the object of a future and more refined treatment of the topic. 

As argued in the introduction, the vacuum manifold for the model of our interest is not a manifold. To define the dynamics on this space we are thus forced to introduce some prescriptions to deal with this pathological issue. This problem could be removed by properly defining a regularization of this space, and more generally it could be of some interest to study the (potential) dependence of the results on the choosen prescription. 

The main motivation of this work was the definition of a
very simple benchmark model for a non-Abelian cosmic string network. In 
particular we tried to define a model that could be evolved with a low request 
of computational power. Most of the networks discussed in this work appear to 
reach a scaling solution that does not make them incompatible with experimental 
observations of our universe. Notice that in the case studied in this work, strings are only loosing energy by transferring it to the scalar field. In order to produce a more realistic description it will be interesting to study a model which allows for the emissions of GW~\cite{DamourVil2,Siemens2}.
In order to produce a more realistic description and to have a deeper understanding of the dynamics it could be interesting to produce a simulation of the same model in a string-oriented simulation~\cite{McGraw1,McGraw:1996py}. In particular in this simulation it would be possible to implement the coupling between the strings and the gravitational field.

\section*{ Acknowledgements}
We would like to thank M. Hindmarsh for helpful suggestions and M.P. would like to thank P. Binetruy and J. Mabillard for useful discussions. We acknowledge the financial support of the UnivEarthS Labex program
at Sorbonne Paris Cit\'e (ANR-10-LABX-0023 and ANR-11-IDEX-0005-02).



\bibliographystyle{JHEP}
\bibliography{biblio}




\end{document}